# High pressure-temperature proton migration in P$\bar{3}$ brucite [Mg(OH)$_2$]: Implication for electrical conductivity in deep mantle

Sudip Kumar Mondal[1], Pratik Kumar Das[2], Nibir Mandal[3]

[1]Department of Physics, Jadavpur University, Kolkata- 700032, India.

[2]Department of Earth Sciences, IIEST, Shibpur, Howrah-711103, India.

[3]Faculty of Sciences, Jadavpur University, Kolkata- 700032, India.

Hydrous minerals contribute largely to the transport and distribution of water into the mantle of earth to regulate the process of deep-water cycle. Brucite is one of the simplest layered dense hydrous mineral belonging to MgO-SiO2-H2O ternary system, which contains significant amount of water in the form of OH- groups, spanning a wide range of pressure stability. Simultaneously, the pressure (p) and temperature (T) induced mobility of protons within the layered structure of brucite is crucial for consequences on electrical conductivity of the mantle. Using ab initio molecular dynamics (AIMD) simulations, we investigate the diffusion of H in high-pressure trigonal P-3 polymorph of brucite in a combined p-T range of 10-85 GPa and 1250-2000K, relevant to the mantle of earth. The AIMD simulations reveal an unusual pressure-dependence of the proton migration in brucite characterized by maximum H-diffusion in the pressure range of 72-76 GPa along different isotherms. We predict that in the P-3 brucite the H mobility is onset only when a critical hydrostatic pressure is attained. The onset pressure is observed to drop with increasing temperature. The H-diffusion in brucite phase at elevated p-T takes place in such a manner that the process results in the amorphization of the H-sublattice, without disturbing the Mg- and O-sublattices. This selective amorphization yields a pool of highly mobile protons causing a subsequent increment in the electrical conductivity in P-3 brucite. Our calculated values of conductivity are compared with ex-situ geophysical magnetic satellite data indicating that brucite can be present in larger quantities in the lower mantle than previously observed. This hydroxide phase can occur as segregated patches between the dominant constituents e.g., silicates and oxides of the lower mantle and thus can explain the origin of high electrical conductivity therein.

## 1. Introduction

The presence of small elements and water have significant effect on the mineralogical structure, composition and dynamics[1,2] of the Earth's mantle which manifests itself in terms of varying melting temperature[3,4], elastic properties[5–13], electrical conductivity[14–19], viscosity, diffusional motion of atoms[20–27] in minerals. It is widely accepted that the carrier of hydrogen into deep earth are a batch of hydrous minerals such as dense hydrous mineral silicates (DHMSs)[28], nominally anhydrous minerals (NAMs)[4,29] and δ-AlOOH[30,31]. However, apart from phase D (ideal formula MgSi$_2$H$_2$O$_6$) most of the hydrous minerals are reported to decompose at high pressures corresponding to the cold subducting slabs [32–35]. Electrical conductivities of DHMSs are observed to increase with pressure suggesting a higher mobility of H atoms. This observation indicates that pressure may act as an ally to enhance movement of protons in crystalline mineral phases. Although quantitatively rare in the mantle of the earth, brucite is an architype hydrous and layered mineral of the MgO-SiO$_2$-H$_2$O ternary system (MSH), which is the most rich in its ability to potentially host water and water-derived species in the mantle[36,37]. In ambient condition brucite assumes a trigonal crystalline structure (space group: P$\bar{3}$m1) where Mg$^{2+}$ and OH$^-$ are arranged in layer. Pressure induced proton frustration in P$\bar{3}$m1 Brucite have been investigated by Raugei et al. [38] and Mookherjee and Strixwood[39]. The former experimental study has showed that under elevated pressure H moves in the *ab* plane and localize separately at three equivalent positions contrary to one in low pressure condition. However, at around 1 GPa brucite undergoes a structural transition to a lesser symmetric trigonal structure (space group: P$\bar{3}$)[40] leading to a change of dynamical positional disorder of proton to a static one[41]. The layered structure of P$\bar{3}$m1 brucite have motivated several researchers to study the diffusion of proton, resulting electrical conductivities[37,42] and it's dehydration properties[43] whereas the proton diffusion in P$\bar{3}$ brucite still remains unexplored. Static DFT calculation and novel structure searching method have demonstrated that P$\bar{3}$ brucite has a larger p-T stability field compared to its low pressure predecessor and around 19 GPa it transforms to a new tetragonal P4$_1$2$_1$2 phase[41]. Nevertheless, this new phase is yet to be experimentally verified. Hermann and Mookherjee [41] also reported that P$\bar{3}$ brucite decomposes into MgO + H$_2$O (liquid) and MgO + ice VII mixtures at high p- low T and high p-high T conditions respectively. But barring the ex-situ geophysical survey of Kirby et al. [44] and subduction zone thermal models of Bina et al. [45], the presence of ice VII phase in the deep interior of the earth remains highly debated. In brucite, high pressure x-ray diffraction study of Fei et al. [46] reveals a smooth diffraction pattern, however in the same paper the authors deduced that at room temperature brucite would decompose into periclase (MgO) and water at 27 GPa. The apparently contradictory results indicate that the decomposition is associated with a high kinetic barrier and brucite is likely to be stable at even higher pressures. Pressure induced enhancement of proton transport in P$\bar{3}$m1 brucite have been reported by Guo and Yoshino[37] but limited to 13 GPa corresponding to the upper mantle regime. Recently, Shaack et al. [47] has demonstrated that in P$\bar{3}$ brucite nuclear quantum effects play a major role in the mobility of H and it reaches a maximum at 67 GPa in room temperature. For the most part, an exhaustive account of the H-diffusion in brucite at high pressure combined with high temperature and its implication for the deep earth is still lacking in literature. On the other hand, electrical conductivities of DHMS and nominally anhydrous minerals

(NAMs) like olivine and its high-pressure polymorph ringwoodite and wadsleyite cannot account for the high conductivity zones in lower mantle. To explain this difference, there should be some other minerals or mineral aggregate which can notably contribute to the electrical conductivity of deep earth. It is important to note that the $P\bar{3}$ brucite structure consists of hollow 2-D well parallel to the ab-plane which are devoid of Mg and O atoms. This channel can offer significant passage for the unhindered motion of H atoms and thereby enhance the electrical conductivity of the mantle.

This study uses Ab-initio Molecular Dynamics (AIMD) to systematically explore the diffusion of H in $P\bar{3}$ brucite in the range 10-80 GPa and 1250-2000K. The diffusion co-efficient of H is calculated in the chosen *p-T* conditions which reveal an anomalous relation of the diffusion coefficients with pressure along the isotherms. The reasons for the differences in diffusivities of H are elucidated. The results demonstrate that the onset temperature of H-diffusion in brucite is largely influenced by the confining pressure, and also deals with the characteristic anisotropy in H diffusion in brucite. Finally, the proton induced electrical conductivity is calculated and compared with deep mantle electrical conductivity to find how far the presence of brucite can affect it.

## 2. Computational methodology

We have performed static, density-functional theory calculations to obtain structures of $P\bar{3}$ brucite at desired pressures up to 80 GPa using the Vienna ab initio simulation package (VASP)[48,49]. The local structural relaxations calculations were performed using the generalized gradient approximation of the Perdew-Becke-Ernzhoff [50] formalism to model electronic exchange-correlation effects, together with projector-augmented wave (PAW) [51] implementation. The PAW-GGA potentials are used to describe the ionic core of H, O and Mg where their valence electronic configurations were $1s^1$, $2s^22p^4$ and $3s^23p^0$, respectively, with core cut-off radii of 1.1, 1.52 and 2.0 Å. The simulations employ an appropriate regular gamma centric 5×5×5 grid of Monkhorst-Pack[52] k points to sample the electron density in the reciprocal space and a kinetic energy cut-off of 625 eV along with a cut-off of 610 eV for the augmented part. These parameters ensured that the precision of the energy calculation is typically lower than 1meV/atom in energy and better than 0.5 GPa in pressure.

To capture the effect of temperature on the motion of atoms, we have performed AIMD simulations as implemented in VASP. The isokinetic NVT ensembles are chosen at a given temperature, T, keeping a fixed volume, V, and the number of atoms, N, in the simulation box. The simulations were performed at volumes corresponding to the desired pressure. The ionic temperatures during the simulations were kept steady employing a Noose-Hoover thermostat[53,54]. Each AIMD simulation was run for 8000-10000 timesteps with each timestep being equal to 1 femto-second, resulting to a total simulation time ranging from 8 to 10 pico seconds. AIMD simulation are typically sensitive of the size of the system under study. To account for this size effect, all the AIMD simulations were carried out on a 2×2×3 supercell containing 180 atoms, obtained from the fully relaxed conventional unit cells from the previous static DFT calculations. The fixed gamma point was used to sample the reciprocal space of the supercells. The constraint imposed by fixed volume along with increasing temperatures affect the resultant pressure. This study considers the thermal pressure correction and, observed them to be within 1.3-2.8% of the target pressure. The images of the crystal structures are rendered using VESTA and the images of the trajectories of H-atoms are extracted from Visual Molecular Dynamics (VMD) suite.

## 3. Result and Discussion

### 3.1 Crystal structure and equation of state

The crystal structure of $P\bar{3}$ brucite consists of $Mg^{2+}$ cations and $OH^-$ anions arranged in layers, assuming an overall trigonal structure. The protons are located in channels in between the edge sharing $MgO_6$ octahedra (Figure 1). The transition from $P\bar{3}m1$ to $P\bar{3}$ brucite, the latter being a maximal subgroup of the former, is characterised by the reduction of mirror planes owing to the spatial disorder of proton distribution. In the lattice structure, the Mg atoms occupy two distinct crystallographic positions: Wyckoff site 1a (0, 0, 0) and 2d (1/3, 2/3, $z_{Mg}$), whereas the O and H atoms occupy general Wyckoff sites 6g and 6i respectively. The most interesting characteristics of the $P\bar{3}$ phase is the occurrence of two different H-H distances[39], where both of them decrease on increasing pressure up to a certain threshold value, beyond which one of them starts to stretch as a consequence of increasing O-H---O angle. Under ambient conditions, the lattice parameters are calculated as $a = b = 5.48529$ Å, $c = 4.79506$ Å and $\alpha = \beta = 90°$, $\gamma = 120°$; these calculated values are in close agreement with previous findings [39,41]. The weak interaction between $MgO_6$ layers is responsible for higher compressibility of this phase along its *c*-axis. The 2nd order Birch-Murnaghan equation of state fit (Supplementary Figure S1) yields a bulk modulus of 46.25 (± 1.7) GPa with a pressure derivative of 5.03, which agree well with the DFT results of Hermann and Mukherjee [41]. The calculated equilibrium volume per formula unit for this phase, 42.02 Å$^3$ is also consistent with their finding.

### 3.2 Proton transport mechanisms

Pressure induced proton frustration and its effect on the mobility of H in brucite has been previously studied both computationally[38] and experimentally [39,55,56] albeit limited to the $P\bar{3}m1$ phase. Quench experiments are also performed to elucidate the phase stability of $P\bar{3}m1$ brucite[40]. Despite those previous studies, the migratory behaviour of H in high pressure P$\bar{3}$ structure and its effects on the electrical conductivities in deep earth remains almost unattended. We have performed AIMD simulations to investigate the kinematic behaviour of proton in P$\bar{3}$ brucite at pressures corresponding to lower and upper mantle. The diffusion constants are calculated from the slope of the mean squared displacement (MSD) versus time curve. It is important to emphasise here that the calculated MSD demonstrates different line segments of varying slope (Figure 2 and Supplementary Figures S2-S5). In order to minimize the error, the final slope was calculated as an average of the slopes of the MSD in several non-intersecting time intervals. Figure S2 shows our calculated MSD of hydrogen atoms at several pressures at the temperature of 1250K, where no notable movement of protons were observed below 43 GPa, which decreases to 28 GPa when the temperature is increased to 1500K as evident from Figure 2.

The onset pressure for proton mobility decreases further to 10.1 and 13.4 GPa when the temperatures are set to 1750K and 2000K, respectively (Supplementary Figure S3-S5). However, irrespective of the temperature to which crystalline brucite is subjected to, the maximum movement of H-atoms are observed in the range 73.6-75.6 GPa. Thus, the MSD's of hydrogen are observed to display an anomalous correlation with confining pressure, decreasing in magnitude upon further increment in pressure. The disorder in proton distribution in P$\bar{3}$ brucite can be categorized into two distinct non-exclusive type: a. dynamic disorder, in which each hydrogen jumps from one to another of the three symmetrically equivalent sites; b. static disorder, in which each hydrogen atom is stationary, in any of the three symmetrically equivalent position[39]. With increasing pressure this disorder changes from a dominantly dynamic one to a dominantly static one. The OH$^-$ interlayer distances are observed to be much more sensitive to pressure compared to the intralayer OH$^-$ distances. At elevated pressure the interactions between OH$^-$ becomes much stronger and results to the reversal of proton disorder in the hydrogen sublattice. At the same time it brings the interlayer H-atoms close to each other to form H-bonds, which are short-lived and very weak [38] aiding the protons to hop between O atoms among Mg-O layer facing each other.

The proton diffusion mechanism in brucite is a complicated process, influenced by Nuclear Quantum Effects [47], that involves two stages: 1) the dissociation of a covalent O-H bond to form another distinct O-H bond, and 2) the reorientation i.e., the jump of proton from one of the three equivalent 6i sites in order to move from an initial O to the nearest one. Pressure has antipodal effect on these two stages. Rising pressure enhances the Nuclear Quantum Effect and increases the dissociation of O-H bonds. On the other hand, the reorientation process is mostly controlled by temperature where pressure is likely to be inclined to localize the proton in a certain orientation, making this motion unfavourable. The dissociation of O-H bonds creates a quasi 2-D proton layer between adjacent MgO layers. At lower pressures two quasi 2-D layers of H atoms are formed near each Mg-O layer and proton move back and forth between them and also throughout the layers. However, at elevated temperature and at pressure between 73-76 GPa these two layers merge and become indistinguishable. We argue that our MSD's at pressures below 75 GPa represent characteristic back and forth movement of proton between two such layers as well as thermally activated motion of proton in each of these layers. At 73-76 GPa the formation of the one indistinguishable layer, populated with large number of mobile protons, enhances the protonic movement. Below the onset pressure at each temperature only the reorientation occurs and resulting in a back-and-forth motion of H atoms leading to negligible net movement. Although these AIMD calculations do not take into account the NQE on an explicit term, the results are in good agreement with that of Schaack et al.[47]

### 3.3 Proton diffusion coefficients

We have systematically calculated the diffusion coefficient of H ($D_H$) at various pressure and temperature conditions using the well-known Einstein's relation

$$D_H = \frac{MSD}{2nt} \quad (1)$$

Where MSD is the mean squared displacement of H-atoms, $t$ is time, $n = 1, 2, 3$ depending on the dimension of the system under study, which in our case is set to 3. Table 1 lists our calculated $D_H$. As the temperature increases the diffusion coefficients also increase. Notably at each temperature we found $D_H$ to assume maximum value in the pressure range 73-76 GPa pressure. Figure 3 shows the variation of $D_H$ with reference to the mantle pressure conditions and different zones in mantle. Several NAMs and transition zone silicates are experimentally observed to host H in their crystalline lattice as substitutional point defects. Olivine aggregates which are dominant species in Earth's lower mantle house nominal amount of water and at higher temperature the H atoms are found to diffuse through the lattice. Discontinuities in seismic wave velocities establishes that olivine undergoes a transition to wadsleyite at 14 GPa and further to Ringwoodite at 24 GPa. Similar to their low-pressure counterpart these transition zone silicates also demonstrate H-mobility at elevated temperatures. Nevertheless, their calculated $D_H$ are one to two order lower than what we have observed in case of brucite at comparable pressure-temperature conditions. The reason behind these differences in $D_H$ can be attributed to their distinctive crystal structure and the different class of mechanisms at play to promote the H-diffusion process. Olivine and Ringwoodite belong to the category of nesosilicates whereas wadsleyite is a sorosilicate. All the former minerals are characterised by the presence of Mg/Fe octahedra and Si-tetrahedra forming a network like structure. In those silicates H must diffuse through asymmetric channels formed by the cationic polyhedra. In contrast brucite is a layered hydroxide mineral which offers an unhindered motion of protons through the layer between $MgO_6$-octahedra. In addition to that, the H-diffusion mechanism in both classes of minerals have remarkable differences. In silicates a net diffusion of H atoms are realised only when an H atom jumps from one substitutional vacancy to the next one. This whole process is thermally controlled and relies on simultaneous creation-annihilation of vacancies together with probabilistic hopping of H through those vacancies. In contrast, the H-diffusion in brucite is initiated and largely regulated by the pressure induced amorphization of the H-sublattice. This pressure induced amorphization creates a pool of mobile H atoms between adjacent $MgO_6$ octahedral layers, depleted of any Mg or O atom to restrict H-mobility. The combined effect of the structure and mechanism of H-diffusion gives rise to a diffusional free energy barrier ranging from 1.66 eV to 3.14 eV for H in ringwoodite and wadsleyite [27] respectively. In contrast, the dissociative and rotational free energy barriers for H in brucite are in the order of 0.01-0.11 eV and 0.03-0.10 eV at room temperature [47], which are expected to drop to even lower values when the temperature rises. Clearly, this higher migration barrier in silicates makes H-diffusion in them kinetically restricted, energetically less favoured and demands relatively higher temperature to initialize as compared to brucite. The apparent free flow of protons and lower migration barrier are thus responsible for the observed high $D_H$ in brucite phase.

The initial high value of $D_H$ observed in 2000K isotherm in the low- pressure regime can be attributed to the incongruent melting of brucite. Supplementary Figure S6 and S7 shows the variations of MSD's of Mg and O atoms at several pressures along 2000K isotherm. At low pressure of ~13 GPa both the MSD's of Mg and O atoms shoot upwards, together with the MSD of H atom (Supplementary Figure S5), indicating that the entire crystalline structure undergoes a melting at this pressure and temperature. However, at higher pressures the melting point of brucite increases, and the H atoms only occur in a mobile state. Supplementary Figure S8 illustrates the MSD of the comparatively heavy atoms at 1250K. It is important to note that at 1250K both the MSD of Mg and O atoms oscillates around some small value indicating the stretching and shortening of Mg-O bonds as the Mg and O atoms execute thermally activated vibrational motions.

### 3.4 Anisotropy in proton diffusion

The proton diffusion in brucite is highly anisotropic in nature. Figure 4a and 4b demonstrates that movements of almost all of the H-atoms are restricted within the planes parallel to crystallographic *ab*-plane, with hardly any out of plane motion observed. The $MgO_6$-polyhedral network here act as a barrier to restrict motions of H-atoms parallel to *c*-axis. I have calculated the axis decomposed diffusion coefficient along the three axes viz. $D_{[100]}$, $D_{[010]}$ and $D_{[001]}$. Further, they are normalized with respect to $D_H$ at corresponding p-T conditions as $d^*_{[100]} = D_{[100]}/D_H$, $d^*_{[010]} =$ $D_{[010]}/D_H$ and $d^*_{[001]} = D_{[001]}/D_H$, respectively. $d^*_{[001]}$ is found to be negligible in magnitude, asserting that proton diffusion along c-axis contributes almost null. The plot of $d^*_{[100]}/d^*_{[010]}$ ratio as a function of pressure in the range 30-90 GPa shows no obvious correlation between the $d^*_{[100]}/d^*_{[010]}$ ratio with pressure-temperature (Figure 4c). Comparable values of $d^*_{[100]}$ and $d^*_{[010]}$ are only obtained at some specific p-T eg. 40GPa-1500K, 70 GPa-1250K, 80 GPa-1500K and 80 GPa-1750K. At those p-T points the movements of H along the *a*- and *b*-axis are equal in magnitude. $d^*_{[100]}/d^*_{[010]}$ at 2000K exhibits maximum anisotropy in diffusion on and above 70 GPa, indicating that protons are much more prone to move along the *a*-axis rather than *b*-axis. Figure 1 shows that the distribution of H-atoms in between $MgO_6$ layers are identical when viewed along *a*- or *b*-axis. When the pressure induced proton disorder and amorphization of the H-sublattice set in, both *a*- and *b*-direction becomes relatively less indistinguishable in terms of proton mobility. The H-atom diffusion parallel to *ab*-plane thus becomes asymmetric in nature without showing any preferred directional dependence.

### 3.5 Analysis of pair distribution functions (PDF)

To explain the unconventional high diffusivity of H in P-3 brucite, we analysed the evolution of pair distribution function (PDF) under varied hydrothermal condition (Fig. 5). Our PDF analysis reveals that over the entire range of pressure and temperature, the H-H PDF (blue line Fig. 5) shows a peak at ~ 1.6 Å between 11-43 GPa, characteristic of the typical nearest neighbour H-H distance in brucite. On further compression the peak moves toward ~1.45-1.50 Å. Notably, the spread of the shoulder in the H-H PDF at higher distances and the lack of shape therein except that of the first peak, indicates the pressure induced amorphization of the H-sublattice. In contrast, the O-Mg PDF (green line Fig. 5) demonstrate a first peak around 1.8-2.0 Å commensurate with the typical O-Mg bond lengths [55,56] and the contraction in those lengths due to increasing external pressure on brucite. The second peak located around ~ 3.3-3.65 Å represents the second nearest neighbour O-Mg distances. It is important to emphasize that both the first and second peaks appearing in the O-Mg PDF are sharp. These sharp peaks are indicative of the fact that even at higher pressure and temperature conditions, under which the H sublattice promptly amorphizes, the Mg-O sublattice still retains it crystalline form i.e., the underlying layered structure of the Mg-O remains unaltered at the onset of the fast proton conduction. Additionally, the increment of pressure narrows and sharpens the first O-Mg peak which suggests that the temperature induced vibrational motions of the O and Mg atoms are restricted to a higher degree.

The O-H PDF (orange line Fig. 5) consists of further intriguing details and shows atypical variation with pressure at different temperatures. At a lower pressure of 11GPa, the first peak in the O-H PDF is the sharpest and very high in amplitude which occurs at ~ 0.98 Å consistent with the O-H bond length in P-3 brucite at that pressure [56]. The right shoulder of the O-H PDF drops to zero and rises to another peak around 2.5 Å [55], located near the trio of the second nearest neighbouring O atoms i.e., the O atoms from the next Mg-O layer in the direction of the H-O bond along the c-axis. As the pressure is increased from 11 GPa to 80 GPa, we observed a minute stretching of the O-H bonds from 0.98 Å to 1.02 Å due to the quenching of interlayer distances which stretches the O-H-O angle as observed by Parise et al[56,57]. Between 50-70 GPa, the another set nearest neighbour peak located at 1.6 Å approach the first peak and the interlayer O-H peak appears ~ 2.35 Å. This is the direct effect of the shortening of Mg-O layer distances which enhances the strength of O-H interaction [40]. At low pressures, the H-atoms are well localized as observed by previous experiments [8,55–58]. However, in addition to the amorphization of the H-sublattice, elevated pressure encourages delocalization of H-atom positions as evident from Fig. 5 c-e. O-H PDF in Fig. 5d shows that apart from the pronounced peak at 1 Å, two other relatively lower amplitude peaks appear within 2.5 Å. This pressure induced delocalization of the of H-atoms thus creates more probable positions for the H-atoms to jump to. The increase in stochastic H-jump frequency due to more available jump sites positively influence $D_H$ as observed in the sharp rise in $D_H$ starting from 40 GPa along different isotherms. To elucidate further, increased pressure forces the adjacent O-H layers to coalesce. From observing the abrupt decrease in the O-H peak distances, we infer that the H atoms which required a jump distance of almost 1.5 Å for producing a net H-diffusion at 10-20 GPa range, need only to traverse a distance of 0.5 Å at higher pressure of 40-70 GPa. This observation is also in par with our previous claim that at high pressure the 2-D channel between adjacent Mg-O layers are populated by highly mobile H atoms. Beyond 70 GPa the 1$^{st}$ and 2$^{nd}$ peak in the O-H PDF merges and gives rise to a less sharp peak spread over almost 0.6 Å. At this pressure, the O-H layers come too close to each other to strengthen the O-H bonds [58] and the delocalization of H positions starts to disappear. One possible reason for such behaviour can be the rotation and spatial rearrangements of the O-H bonds taking precedence over the simultaneous creation and annihilation of O-H bonds which reduces the probability of jump of H-atom, eventually resulting in a lower $D_H$ compared to the $D_H$ at 70 GPa.

### 3.5 Electrical conductivity

To estimate the contribution of conductivity of the H-atom, $\sigma_H$, to the apparent bulk conductivity of the system, we have used the alternative form of the Nernst-Einstein equation

$$\sigma_H = \frac{1}{k_B T} D_H z_H^2 e^2 c_H \quad (2)$$

where $Z_H$ and $e$ are the valence of the diffusing species and charge of an electron respectively. $k_B$ is the thermodynamic Boltzman constant, and T is the simulation temperature. $C_H$ is the concentration of H atoms i.e., number of atoms/unit volume. The increased proton diffusion in brucite results in an increment in electrical conductivity (σ) as both the temperature and pressure are raised (Figure 6). Electrical conductivity of dry and wet P$\bar{3}$m1 brucite have been experimentally calculated by Gasc et al. [42] but limited to the pressure of 2 GPa only. At these pressures the wet sample is found to show an electrical conductivity in the range $10^{-2}$ to $10^{-3}$ S/m at 1173K, which is reasonably low compared to our calculations. Guo and Yoshino [37] did a similar study on similar crystalline brucite and observed a maximum conductivity of 32 S/m at pressure 11-13 GPa. This value is comparable to the calculated electrical conductivity at *p-T* points of 28GPa-1500K, 10GPa-1750K and 18.7 GPa-2000K respectively. Even much higher values of σ are observed experimentally in DHMS phase A, phase D and the super-hydrous phase B by Guo and Yoshino [59]. Phase A features a σ of 55 S/m at 10 GPa in the temperature range 500-900K, phase D on the other hand shows an electrical conductivity of 1342 S/m at 22 GPa in the same temperature range. We have obtained comparable values of the electrical conductivities in P$\bar{3}$ brucite but only in the pressure range 50-60 GPa and between temperature 1500-2000K. This asserts that such high conductivities are not unusual. In fact, the removal of the mirror plane and lowering of symmetry in pressure induced P$\bar{3}$m1 to P$\bar{3}$ transition in brucite allows more space for H to diffuse rapidly. At pressures higher than 60 GPa, our calculated values of σ surpass the conductivities of DHMS phases. Although the diffusion of H is characterised by a maximum value in the range 73-76 GPa, for 1250k and 2000K we observed that the maximum σ is attained beyond this pressure range despite the lowering of diffusion coefficients. For the other two temperatures the trend of the variation of σ is qualitatively similar to what we observed for $D_H$ at those temperatures.

In Figure 6 we have compared our calculated σ values with mantle electrical conductivity using magnetic satellite measurements by Constable and Constable[60]. At low pressure regime our σ values are in good agreement with their data. Their observation shows a seemingly rapid increment of σ at around 50 GPa, however this study is limited to 60 GPa in pressure. The calculated σ also demonstrates a rapid increment in similar range of pressure although they don't converge to a fixed value as the pressure in increased further. Our observation together with the identification of high σ in DHMS by Guo and Yoshino [59] indicates the presence of proton disordered brucite in lower and upper mantle region of the earth. The mean electrical conductivity of the mantle ranges from $10^{-4}$ to $10^3$ S/m[61] which is lower than our calculated conductivities of brucite beyond 60 GPa corresponding to the upper mantle. Thus, this study infers that the amount of brucite in shallow mantle could be moderate to high but reasonably small in lower mantle and brucite could occur as independent pockets at those depths.

### 3.6 Conclusion

This study has systematically investigated the proton diffusion behaviour in P$\bar{3}$ brucite at high pressure and high temperature regime. The study reveals an anomalous behaviour of hydrogen diffusion where the diffusion constants increase up to a certain pressure and then exhibit maxima in 73-76 GPa pressure range across all the isotherms. At this pressure range two separate layers of protons between MgO$_6$ octahedral sheets emerge and coalesce with each other. This coalition of proton layers generates high number of free protons. At high temperature the hydrogen sublattice amorphized leaving the Mg and O atoms static in their lattice sites. The degree of amorphization increases with increasing temperature and thus yields highly mobile protons. Beyond this pressure, the coalition of proton layers becomes ineffective and thereby reduces the diffusion constant. The arrangement of H in layered structure of P$\bar{3}$ brucite is identical along crystallographic a- and b-axis. The calculated anisotropy in proton migration thus reveals no axial preference but indicates towards the random thermal motion of protons, other than the fact that no net diffusion of proton is observed along c-axis, which was present in P$\bar{3}$m1 brucite[37].

AIMD calculations are used to evaluate the apparent contribution of the protonic conductivities to the electrical conductivities of brucite under varied hydro-thermal conditions. While the diffusion constants are observed to increase steadily with temperature, the electrical conductivities offer a complex variation. For 1500K and 1750K, the maximum of conductivity coincides with the same p-T points where the diffusion constant shows maxima, whereas for 1250K and 2000K the conductivities are observed to increase further with temperature. At pressures corresponding to upper mantle the conductivity features very high values comparable to several DHMS phase. Comparison with geomagnetic data [60,61] allows us to conclude that apart from predominant constituents of the mantle such as silicates and oxides, brucite can also be present in the earth's mantle in small amount.

# List of figures with captions

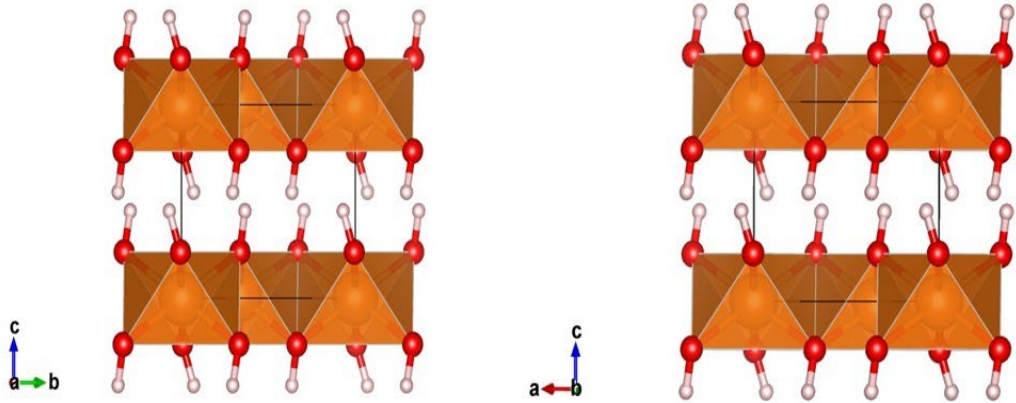

**Figure 1:** Crystal structure of $P\bar{3}$ brucite as seen from perpendicular to a-axis (left) and b-axis (right). The brown polyhedra are MgO$_6$ octahedra. Red and Pink spheres are O and H atoms, respectively. Note the 2D hollow channel parallel to ab-plane.

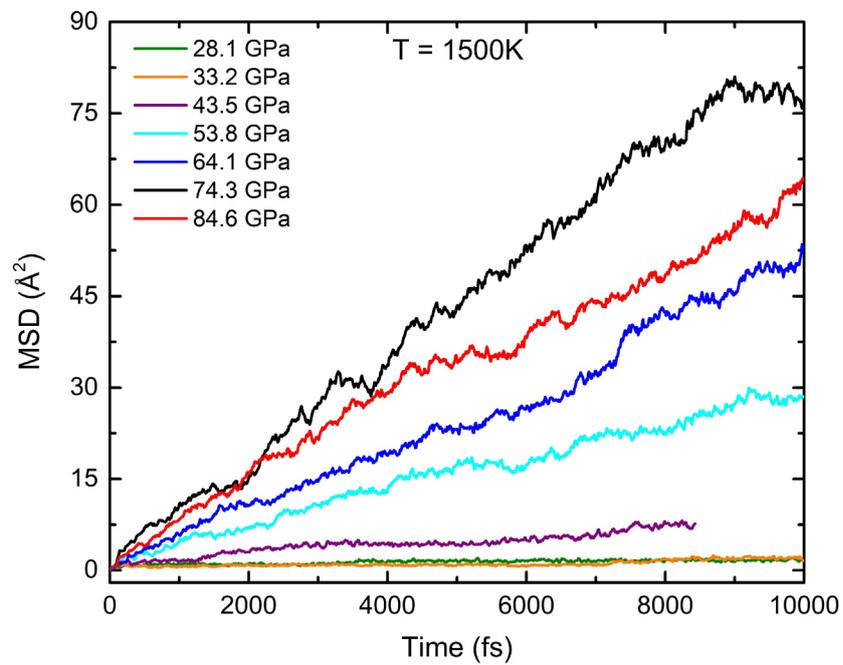

**Figure 2:** Mean square displacements of hydrogen atoms at 1500K. The maximum mobility of H is observed at 74.3 GPa.

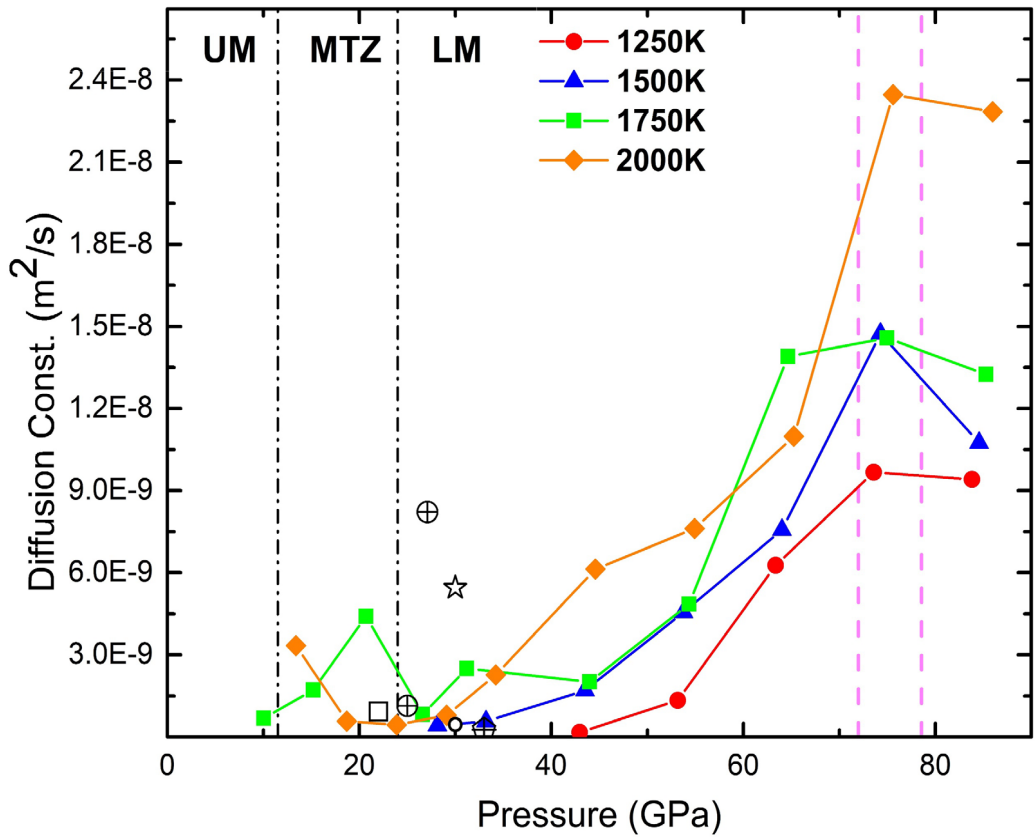

**Figure 3:** Proton diffusion co-efficient at various p-T condition. LM- Lower Mantle, UM- Upper Mantle, MTZ- Mantle Transition zone. The black dash-dotted vertical lines at 11 and 24 GPa represents the boundary between LM and MTZ and between MTZ and UM respectively. The zone between the magenta vertical lines represents the high-pressure zone where maximum H-diffusion occurs. For comparison the data of H-diffusion in transition zone silicates i.e., in ringwoodite and wadsleyite after Caracas and Panero [27] are also included. Ringwoodite: circle with + sign- $Mg^{2+} \rightarrow 2H^+$ at 2000K and 2500K, star- $Si^{4+} \rightarrow Mg^{2+} + 2H^+$ at 2500K, open circle- both $Mg^{2+} \rightarrow 2H^+$ and $Si^{4+} \rightarrow Mg^{2+} + 2H^+$ at 2500K, pentagon - $Si^{4+} \rightarrow 4H^+$ at 2500K; wadsleyite: open square- $Mg^{2+} \rightarrow 2H^+$ at 2000K. Note the characteristic large proton diffusion coefficient in brucite at high pressures.

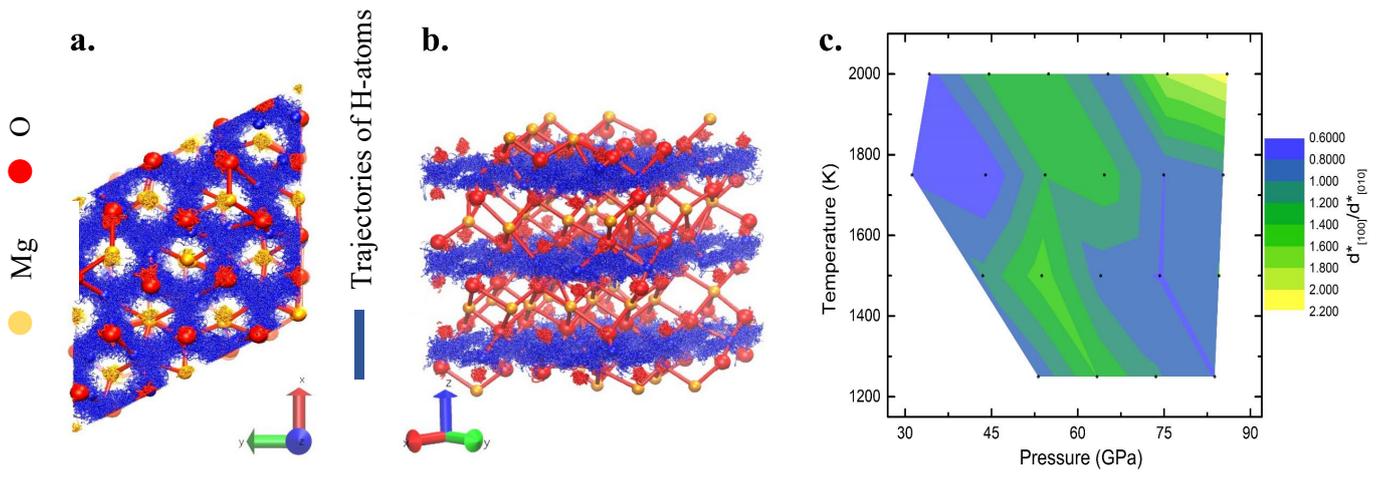

**Figure 4:** Anisotropic character of proton diffusion in brucite. The trajectories of H-atoms (blue lines in a. and b.) are constrained within diffusive layers of H-atoms sandwiched between MgO$_6$ octahedral layers.

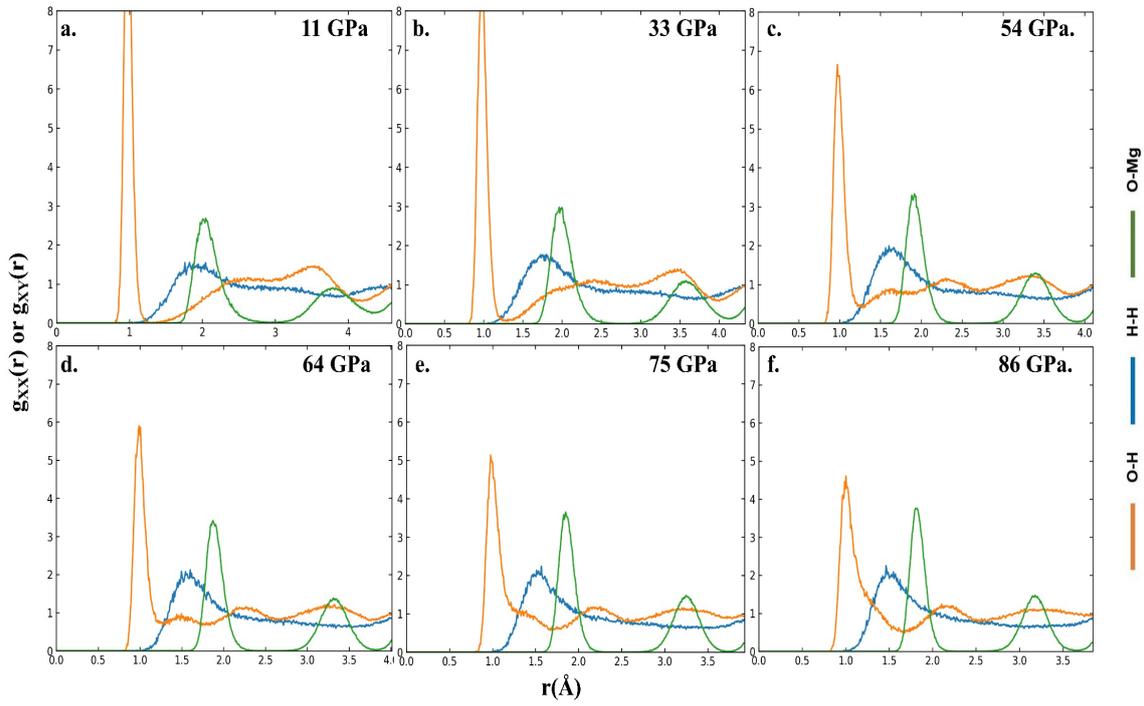

**Figure 5:** Analysis of Pair Distribution Function (PDF) of P-3 brucite at 1500K and different pressures. The emergence of new peaks around the central O-H peak suggest pressure induced delocalization of H-atoms with respect to O atoms, whereas the loss of long-range order in H-H PDF suggests amorphization of H-sublattice.

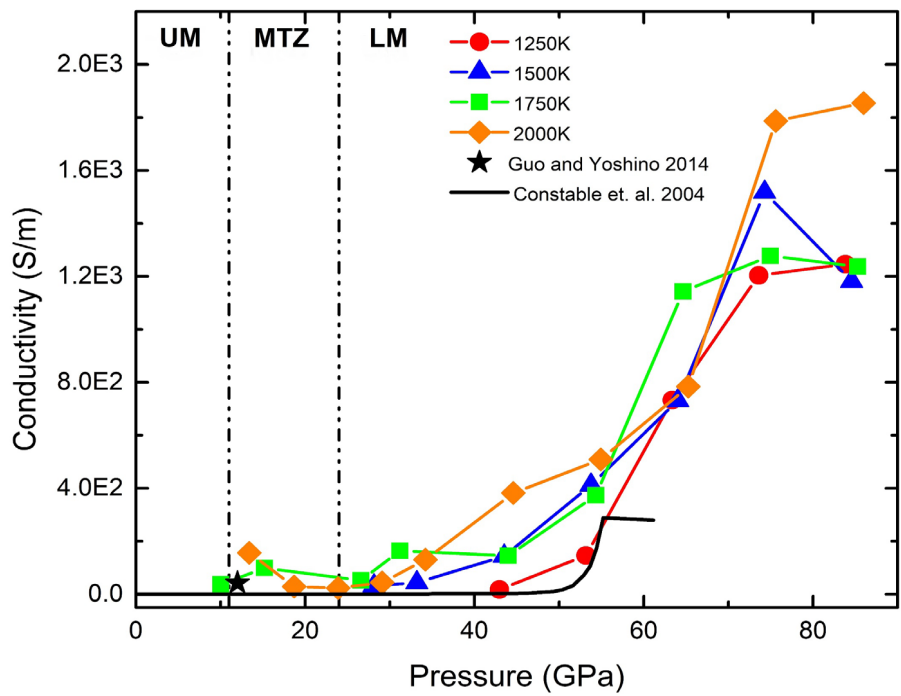

**Figure 6:** Variations of protonic conductivity in P$\bar{3}$ brucite as a function of pressure along different isotherms. The black star represents electrical conductivity of P$\bar{3}$m1 brucite at 11-13 GPa range [ref. 37].

## List of table(s) with caption(s):

**Table 1:** Calculated diffusion coefficients of H in brucite at temperature 1250-2000K and pressures 10-85 GPa. The pressure includes the thermal correction as well.

| Temperature (K) | Pressure (GPa) | Diffusion Coefficient $D_H$ (m²/s) |
|---|---|---|
| 1250 | 42.95 | 1.7011E-10 |
| | 53.16 | 1.3363E-09 |
| | 63.37 | 6.2678E-09 |
| | 73.59 | 9.6683E-09 |
| | 83.82 | 9.4072E-09 |
| 1500 | 28.13 | 4.20345E-10 |
| | 33.19 | 5.65935E-10 |
| | 43.51 | 1.69786E-09 |
| | 53.76 | 4.55502E-09 |
| | 64.02 | 7.56022E-09 |
| | 74.28 | 1.47315E-08 |
| | 84.57 | 1.07429E-08 |
| 1750 | 10.04 | 6.79615E-10 |
| | 15.26 | 1.71247E-09 |
| | 21.32 | 4.40833E-09 |
| | 26.61 | 8.18315E-10 |
| | 31.20 | 2.50285E-09 |
| | 44.0 | 2.01455E-09 |
| | 54.35 | 4.85519E-09 |

|  | 64.64 | 1.39001E-08 |
|---|---|---|
|  | 74.96 | 1.45728E-08 |
|  | 85.28 | 1.32427E-08 |
| **2000** | 13.4 | 3.3315E-09 |
|  | 18.7 | 5.74253E-10 |
|  | 23.9 | 4.42703E-10 |
|  | 29.09 | 7.94527E-10 |
|  | 34.22 | 2.26957E-09 |
|  | 44.6 | 6.13043E-09 |
|  | 54.93 | 7.61753E-09 |
|  | 65.27 | 1.09825E-08 |
|  | 75.6 | 2.34688E-08 |
|  | 85.99 | 2.28415E-08 |

# Supplementary materials

# Supplementary figures with captions

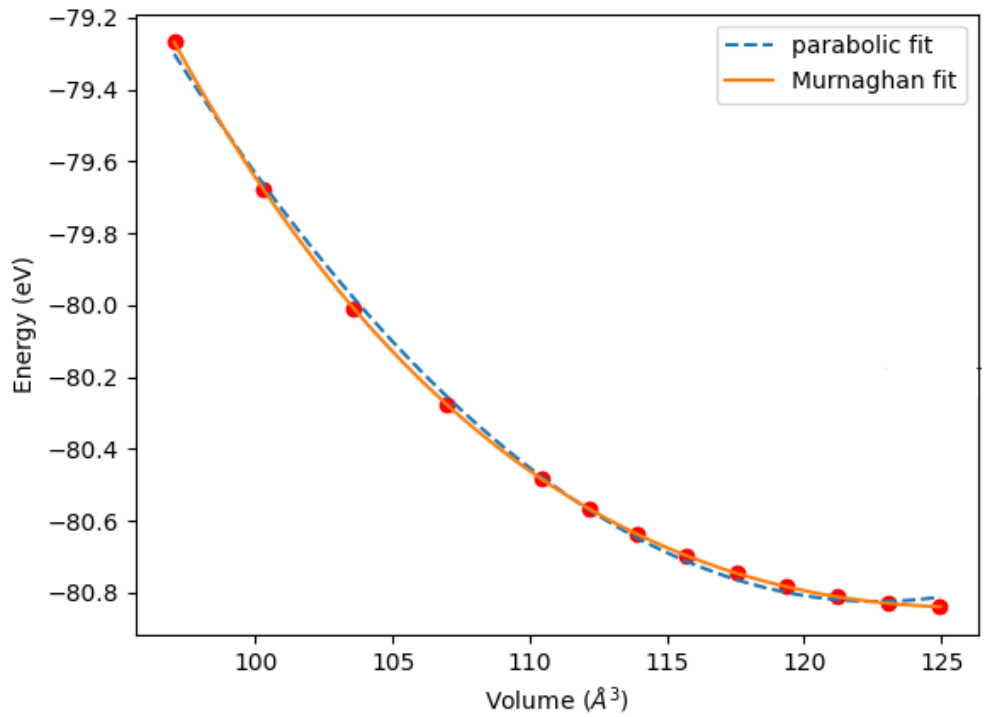

**Figure S1:** Second order Birch-Murnaghan equation of state of P$\bar{3}$ brucite.

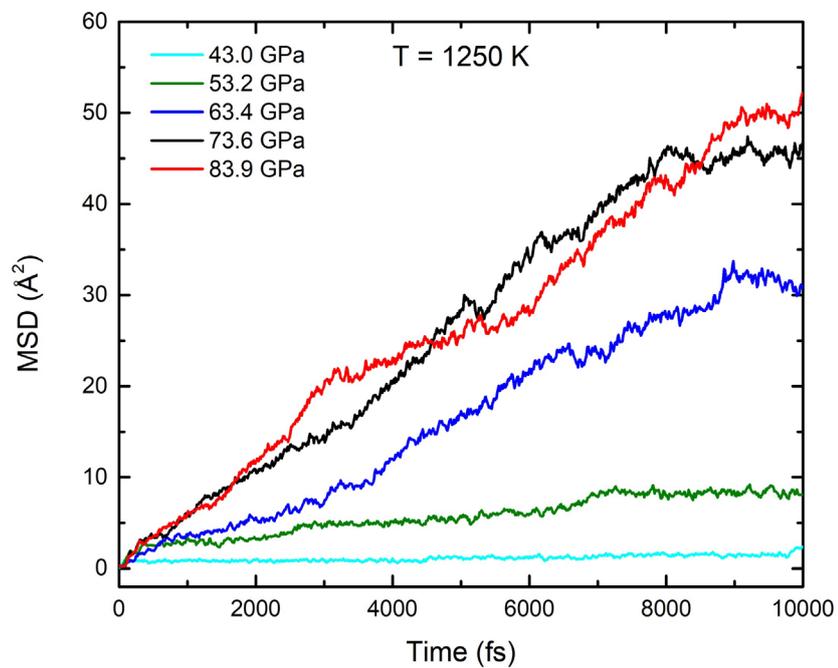

**Figure S2:** Mean squared displacements of H-atom at 1250K. Note that a reasonable proton diffusion is not observed until the pressure rose to a value near 43GPa at this temperature.

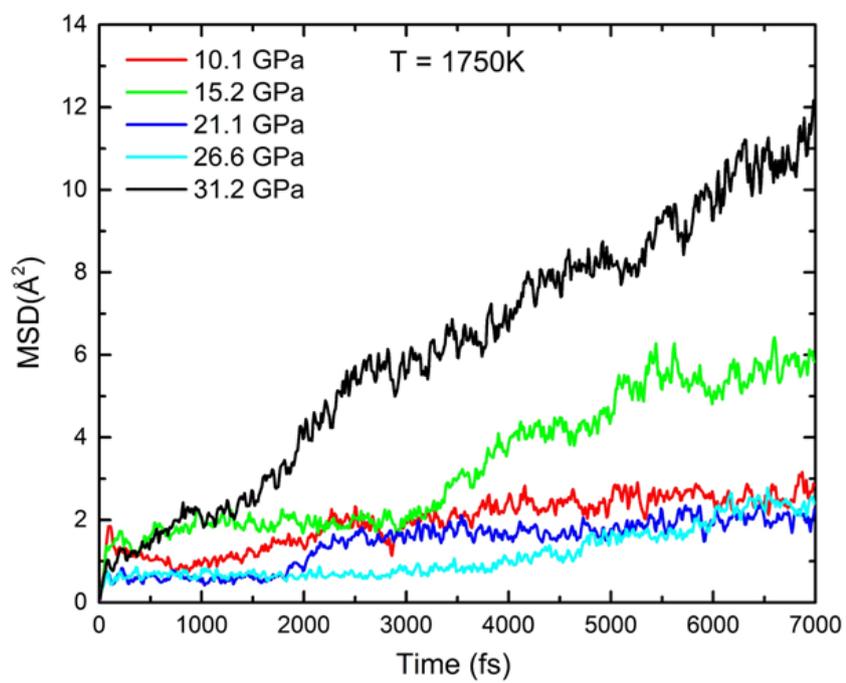

**Figure S3:** MSD of H atoms at 1750K and up to pressure of 31.2 GPa.

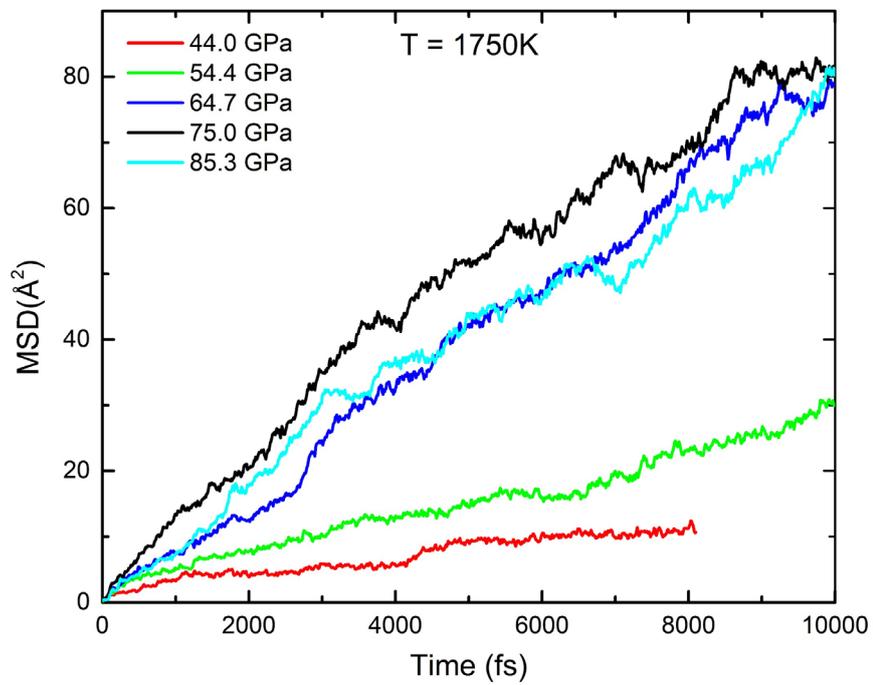

**Figure S4:** MSD of H atoms at 1750K and high pressures. Note the peak MSD at 75.0 GPa.

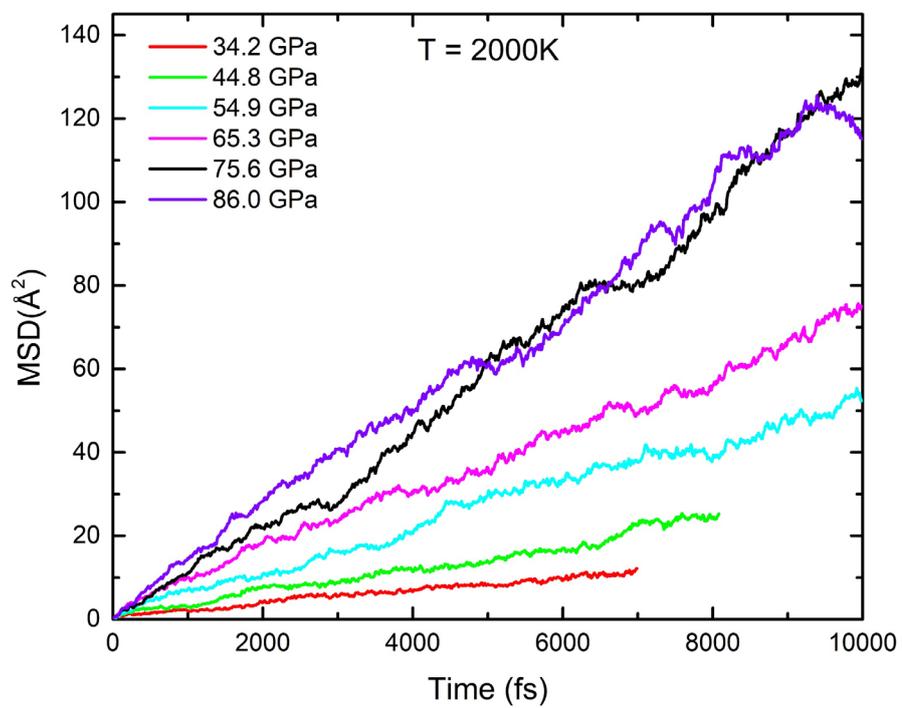

**Figure S5:** MSD of H atoms in brucite at 2000K. For brevity only the MSD's above 30 GPa are shown here.

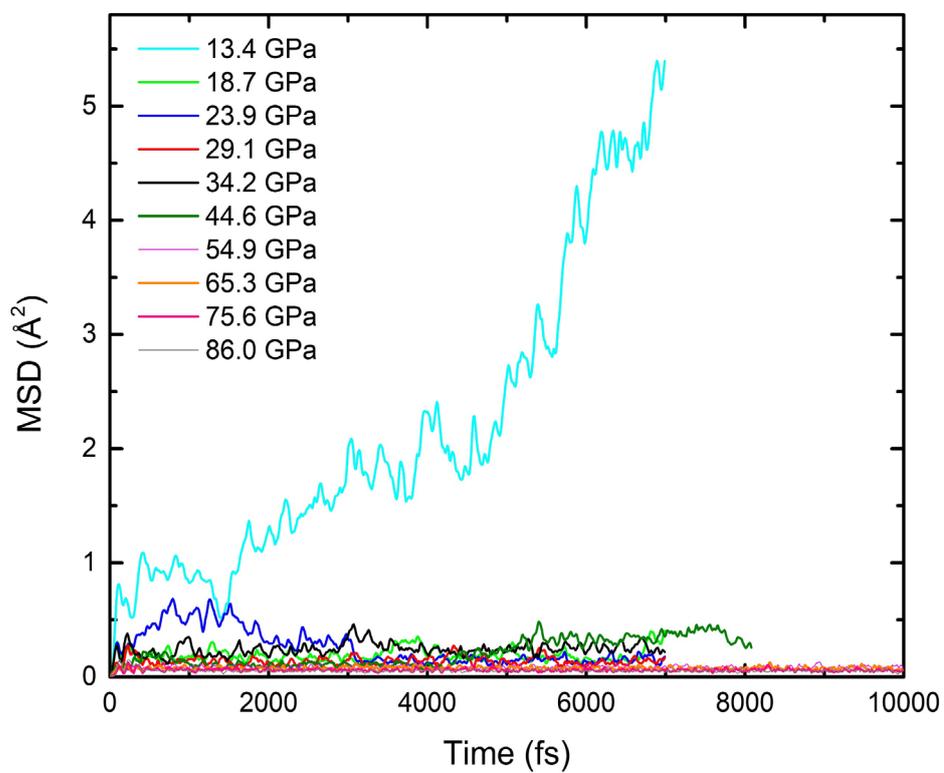

**Figure S6:** MSD of Mg atoms at different pressures while the temperature was kept fixed at 2000K. Note the large MSD of Mg atoms at 13.4 GPa indicating the melting of brucite.

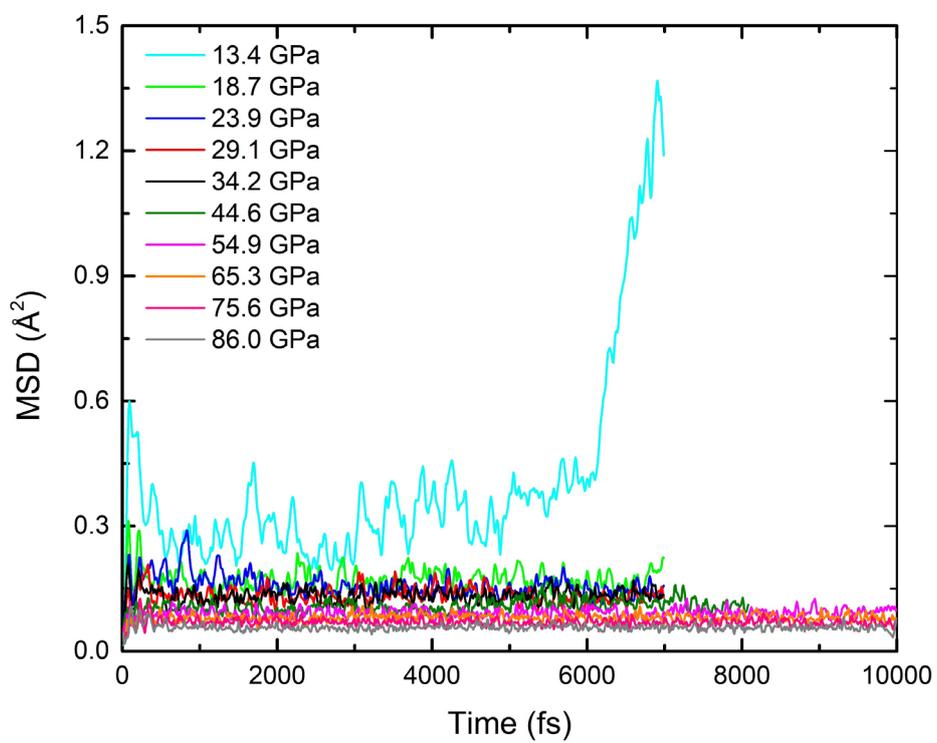

**Figure S7:** MSD of O atoms in brucite at 2000K. The O atom displays a qualitatively similar increment in MSD as the Mg atoms at 13.4 GPa.

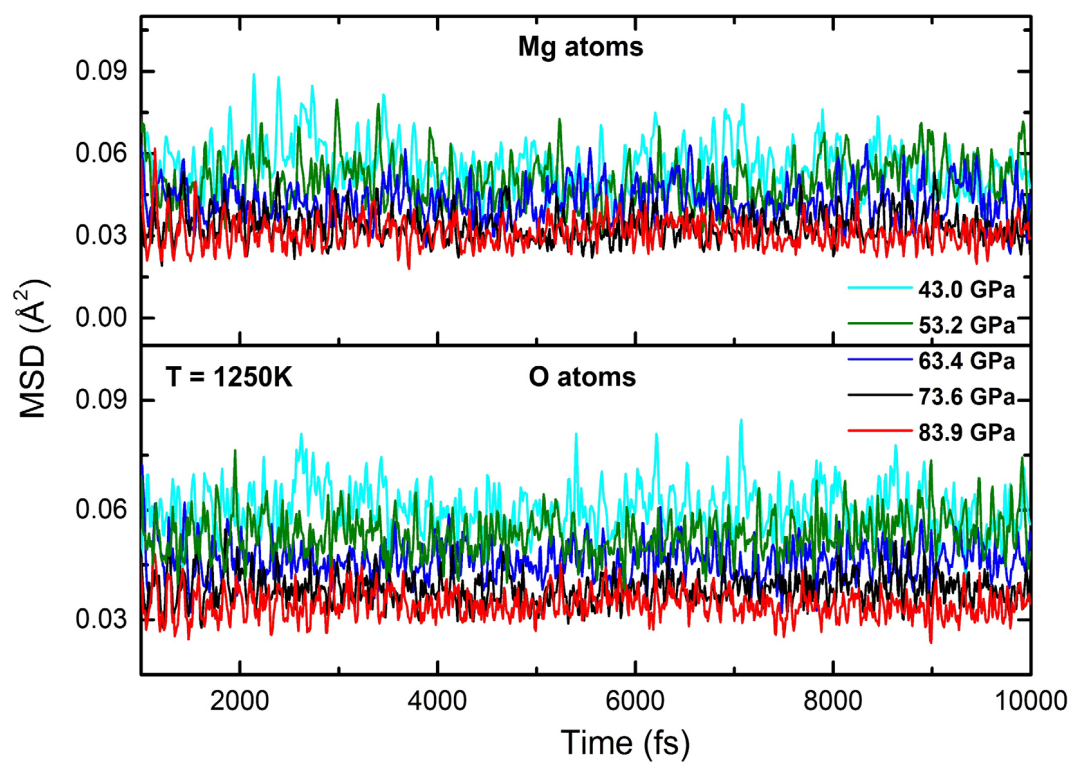

**Figure S8:** MSD's of Mg and O atoms at 1250K. The relatively minute values MSD's suggest that the Mg and O atoms are executing vibrational motions about their mean positions.